\documentclass[preprint,fleqn,pre,twocolumn,10pt,showpacs,balancelastpage]{revtex4}
\usepackage{amsmath}
\usepackage{graphicx}

\input{tcilatex}

\begin{document}

\title{Self--pulsing dynamics of ultrasound in a magnetoacoustic resonator}
\author{V. J. S\'{a}nchez--Morcillo, J. Redondo, J. Mart\'{\i}nez--Mora, V.
Espinosa and F. Camarena}
\affiliation{Departament de F\'{\i}sica Aplicada, Universitat Polit\`{e}cnica de Val\`{e}%
ncia, Crta. Natzaret-Oliva s/n, 46730 Grau de Gandia, Spain}

\begin{abstract}
A theoretical model of parametric magnetostrictive generator of ultrasound
is considered, taking into account magnetic and magnetoacoustic
nonlinearities. The stability and temporal dynamics of the system is
analized with standard techniques revealing that, for a given set of
parameters, the model presents a homoclinic or saddle--loop bifurcation,
which predicts that the ultrasound is emitted in the form of pulses or
spikes with arbitralily low frequency.
\end{abstract}

\maketitle

\section{Introduction}

Magnetostriction is essentally a nonlinear phenomenon, which accounts for
the change of dimension of a magnetic material under applied magnetic
fields. However, most of the applications of this phenomenon consider small
amplitude fields and consequently exploit only their linear properties, as
is the case of electromagnetic-acoustic transducers. The consideration of
nonlinearities in the description of this process lead to the discovery of
new properties, such as the parametric wave phase conjugation (WPC), a
phenomena which is actually an active field of research with promising
applications into acoustic microscopy \cite{Brysev00} or harmonic imaging 
\cite{Krutyansky02}. A review on WPC theory and methods can be found in \cite%
{Brysev98}. There is also increasing interest in the developement of
magnetostrictive transducers working at high powers, where nonlinear effects
are not negligible. The advances in this field come in paralel with the
search of novel magnetic materials with high magnetostriction values.

On the other side, parametric phenomena in different fields of nonlinear
science share many common features, such as bistability, self-pulsations and
chaos among others. Is precisely this analogy that motivates our search for
complex dynamical phenomena in parametrically driven magnetoacoustic systems.

Different models have been proposed for the description of magnetoacoustic
interaction in ferromagnetic materials. Also, many experimental progress in
this field has been achieved, and the main parameters involved in the
process have been determined \cite{Brysev200}. In this paper we analize the
dynamical behaviour of this system in the presence of magnetic nonlinearity.
It is shown that a cubic magnetic nonlinearity is responsible for the
appearance of new effects not reported before in this system, such as
self-pulsing dynamics and spiking behaviour, related to the existence of
homoclinic bifurcations

\section{The model}

The physical system considered in this paper consists in a electric $RLC$
circuit, driven by an external ac source at frequency $2\omega $ and
variable amplitude $\varepsilon $. The inductance coil, with density of
turns $n$, transverse section $S$ and length $L,$ contains a ferromagnetic
ceramic material which acts as an acoustical resonator and is the origin of
nonlinearities in the system when the driving $\varepsilon $ is high enough.
A theoretical model for this system has been derived in the resonant case in 
\cite{Streltsov97}. We review here the details of the derivation for
convenience.

The equation for the circuit is $\varepsilon _{R}+\varepsilon
_{C}+\varepsilon _{L}=\varepsilon _{ext}$, where $\varepsilon _{ext}=%
\mathcal{E}\cos (2\omega t),$ $\varepsilon _{R}=IR,$ $\varepsilon _{C}=q/C$,
and $\varepsilon _{L}$ depends nonlinearly on the magnetic fields.

We consider two effects that introduce the nonlinearities in the problem.
The first source of nonlinearity results from magnetoacoustic interaction,
which appears via phonon-magnon proccesses. In the case of parametric
magnetostriction ceramics, the generation of ultrasound takes place at half
of the frequency of the driving. This process is represented, in general, by
a lagrangian density term in the form $\mathcal{L}_{int}=\alpha
_{ijk}H_{i}u_{j}u_{k}$,\cite{Streltsov97} where $u_{k}$ is the acoustic
displacement component in the direction $k$ ($k=x,y,z$) and $\alpha $ has in
general tensor character.

The total magnetic induction $H$ acting along the axis of the material is
the result of three contributions: a static field $H_{0}$ produced, e.g. by
a permanent magnet surrounding the active ceramic material or a coil
carrying a stationary current, an alternating field $H_{q}\left( t\right) $
induced by the ac current in the circuit, and finally a field produced by
material deformations $H_{int}$, which results from the magnetoacoustic
interaction. From the lagrangian density it follows that, in the case of
longitudinal waves propagating in the $z$ axis ($u_{x}=u_{y}=0)$, the
acousto-induced magnetic field has the form%
\begin{equation}
H_{int}=-\alpha \frac{1}{V}\dint u(\mathbf{r},t)^{2}dV,  \label{Hint}
\end{equation}%
where $\alpha =\alpha _{zzz}$ is the coefficient of coupling between the
active medium of volume $V$ and the pump source. Thus, the effective
magnetic induction takes the form%
\begin{equation}
H=H_{0}+H_{q}(t)+H_{int}(t).  \label{H}
\end{equation}%
where we assume that the relation $H_{0}>>H_{q}(t),H_{int}(t)$ holds.
Alternatively, $H$ given by (\ref{H}) can be considered as a nonlinear
pumping field \cite{Preobrazhensky93}. Note that the appearance of the
subharmonic field saturates the magnetic field value, and modifies the
effective pumping.

A second source of nonlinearity is typical of ferromagnetic materials. We
assume that, for weak saturation, the scalar nonlinear relation between the
magnetic field and the magnetic induction, $B=\mu \left( H\right) H$, can be
written, following \cite{AbdAlla87}, as

\begin{equation}
B=\mu H+\frac{1}{6}\mu _{0}\chi ^{(3)}H^{3},  \label{B}
\end{equation}%
where $\mu $ is linear the permeability of the material and $\chi ^{(3)}$
the third order magnetic susceptibility, which in turn depends on the
frequency.

Applying the Faraday law under the previous assumptions, considering only
resonant terms oscillating at the frequency of the driving $2\omega $, and
neglecting terms higher than quadratic in $H_{q}$ and $H_{int}$, we get 
\begin{eqnarray}
\varepsilon _{L} &=&N\frac{d\Phi _{B}}{dt}=N\frac{d}{dt}\dint \mathbf{B}~d%
\mathbf{s=}\mathcal{L}\frac{dI}{dt}+  \label{eps} \\
&&\mu N\frac{d}{dt}\dint H_{int}~ds+\mu _{0}\chi ^{(3)}N\frac{d}{dt}\dint
H_{0}H_{q}H_{int}~ds,  \notag
\end{eqnarray}%
where $I=dq/dt$ is the electrical current in the circuit, $\mathcal{L}$ is
the inductance and $H_{q}=NI/L.$ In terms of the charge in the capacitor,
and taking into account that $V=SL$ and that $n=N/L$ is the density of
turns, the circuit equation takes the form%
\begin{eqnarray}
&&\mathcal{L}\frac{d^{2}q}{dt^{2}}+R\frac{dq}{dt}+\frac{q}{C}=\mathcal{E}%
\cos (2\omega )+\mu n\alpha \frac{d}{dt}\dint u^{2}~dV+  \notag \\
&&\mu _{0}\chi ^{(3)}n^{2}H_{0}\alpha \frac{d}{dt}\left( \frac{dq}{dt}\dint
u^{2}~dV\right) ,  \label{charge}
\end{eqnarray}%
where the last two terms represent the nonlinearities related magnetoelastic
interaction and magnetic nonlinearity respectively.

Let us consider now the evolution of the acoustic wave. Considering the
current induced magnetic field $H_{q}$ as the main source term, we find \cite%
{Preobrazhensky93} 
\begin{equation}
\frac{1}{v^{2}}\frac{\partial ^{2}u}{\partial t^{2}}-\nabla ^{2}u=\alpha
H_{q}(t)u=\alpha n\frac{dq}{dt}u,  \label{sound}
\end{equation}%
where $v$ is the propagation velocity of sound in the material. Here the
effect of the acousto-induced magnetic field (quadratic in the small
coupling constant $\alpha $) has been neglected. However, we have checked
that the consideration of this small term do not imply qualitative changes
in the results reported below.

We consider solutions of Eqs. (\ref{charge}) and (\ref{sound}) in the form
of quasi-harmonic waves, i.e., whose amplitudes are slowly varying in time.
In this case we can write 
\begin{subequations}
\begin{align}
q\left( t\right) & =\frac{1}{2}\left[ Q\left( t\right) \exp \left( 2i\omega
t\right) +c.c.\right] ,  \label{fieldQ} \\
u\left( \mathbf{r},t\right) & =\frac{1}{2}\left[ U\left( t\right) \exp
\left( i\omega t\right) +c.c.\right] g\left( \mathbf{r}_{\bot }\right) \sin
\left( kz\right) ,  \label{fieldU}
\end{align}%
where $u\left( \mathbf{r},t\right) $ has the form of cavity modes of the
acoustical resonator, $k=m\pi /L$ define the cavity resonances, $2\omega =1/%
\sqrt{\mathcal{L}C}$ is the pumping frequency resonant with the circuit, and 
\end{subequations}
\begin{equation}
\left\vert \frac{dX}{dt}\right\vert <<\left\vert \omega X\right\vert ,
\end{equation}%
with $X=Q$ or $U,$ represents the slowly-varying envelope approximation.

Under these assumptions, the slow amplitudes obey the evolution equations 
\begin{subequations}
\label{eqsQU}
\begin{align}
\frac{dQ}{dt}& =iE-\gamma _{Q}Q+\xi _{1}U^{2}+i\xi _{2}\left\vert
U\right\vert ^{2}Q,  \label{eqsQ} \\
\frac{dU}{dt}& =-\gamma _{U}U+\xi _{3}QU^{\ast },  \label{eqsU}
\end{align}%
where $E=\mathcal{E}/4\omega \mathcal{L}$, and $\gamma _{Q}=R/2\mathcal{L}\ $%
and $\gamma _{U}$ represent the electric and acoustic losses respectively.
The last parameter is introduced phenomenologically, and take into account
the losses due mainly to radiation from the boundaries. The coefficients of
the nonlinear terms are defined as 
\end{subequations}
\begin{align}
\xi _{1}& =\frac{\mu n\alpha L}{8\mathcal{L}}\dint g\left( \mathbf{r}_{\bot
}\right) ^{2}\text{d}\mathbf{r}_{\bot },  \notag \\
\xi _{2}& =\frac{\omega \mu _{0}\chi ^{(3)}n^{2}\alpha H_{0}L}{4\mathcal{L}}%
\dint g\left( \mathbf{r}_{\bot }\right) ^{2}\text{d}\mathbf{r}_{\bot },
\label{coef} \\
\xi _{3}& =\frac{v^{2}n\alpha }{2}.  \notag
\end{align}%
Equations (\ref{eqsQU}) are the model considered in \cite{Streltsov97} with
some corrections. This system can be further simplified using the
normalizations 
\begin{equation}
Q=\left( \frac{\gamma _{U}}{\xi _{3}}\right) X,\ \ U=\sqrt{\frac{\gamma
_{U}\gamma _{Q}}{\xi _{1}\xi _{3}}}Y,\ \ E=\left( \frac{\gamma _{U}\gamma
_{Q}}{\xi _{3}}\right) \mathcal{P},  \label{norm}
\end{equation}%
which leads to the model 
\begin{align}
\frac{dX}{d\tau }& =\mathcal{P}-X+Y^{2}+i\eta \left\vert Y\right\vert ^{2}X,
\label{eqs2} \\
\frac{dY}{d\tau }& =-\gamma \left( Y-XY^{\ast }\right) ,  \notag
\end{align}%
where $\gamma =\gamma _{U}/\gamma _{Q}$ is the ratio between losses, $\tau
=\gamma _{Q}t$ is a dimensionless time and%
\begin{equation}
\eta =\frac{4\omega H_{0}\gamma _{U}\chi ^{(3)}}{\alpha v^{2}\mu _{r}}.
\label{eta}
\end{equation}%
remains as the single parameter of nonlinearity, with $\mu _{r}=\mu /\mu
_{0} $. Due to the number of parameters involved in $\eta $, it can be
varied over a wide range of values. Note that Eqs. (\ref{eqs2}) also possess
the symmetry $Y\rightarrow -Y$.

\section{Stationary solutions and stability}

Equations (\ref{eqs2}) possess two kinds of stationary solutions. For small
pump values, the acoustic subharmonic field is absent, and the homogeneous
trivial solution is readily found as%
\begin{equation}
\left\vert X\right\vert =\mathcal{P},\ \left\vert Y\right\vert =0.
\end{equation}%
For higher values of the pump, the trivial solution becomes unstable and the
acoustic subharmonic field is switched-on. The corresponding values of the
amplitudes are given by 
\begin{equation}
\left\vert X\right\vert =1,\ \ \left\vert Y\right\vert ^{2}=\frac{1\pm \sqrt{%
\mathcal{P}^{2}(1+\eta ^{2})-\eta ^{2}}}{1+\eta ^{2}},  \label{inthom}
\end{equation}%
and the bifurcation occurs at a critical (threshold) pump given by 
\begin{equation}
\mathcal{P}_{th}=1,
\end{equation}%
which is independent of the value of the nonlinearity coefficient $\eta $.

From (\ref{inthom}) it also follows that the acoustic field shows
bistability, i.e., both solutions coexist for pump values between $\mathcal{P%
}_{b}$ and $\mathcal{P}_{th}$, where $\mathcal{P}_{b}$ corresponds to the
turning point of the solution and is given by 
\begin{equation}
\mathcal{P}_{b}=\sqrt{\frac{\eta ^{2}}{1+\eta ^{2}}}.
\end{equation}

We consider next the stability of the homogeneous solutions (\ref{inthom}),
by means of the well known linear stability analysis technique. Substituting
in Eqs. (\ref{eqs2}) and their complex conjugate a perturbed solution in the
form $x_{i}\left( t\right) =\overline{x}_{i}+\delta x_{i}(t),$ where $%
\overline{x}_{i}$ is a vector with the particular stationary values, and
linearizing the resulting equations around the small perturbations, one
obtains that $\delta x_{i}\sim e^{\lambda t}$, where $\lambda $ are the
eigenvalues of the stability matrix that relates the vector of the
perturbations with their temporal derivatives. The instability of the
solution is determined by the existence of positive real parts of the roots $%
\lambda $ of the fourth-order characteristic polynomial%
\begin{equation}
P\left( \lambda \right) =\lambda ^{4}+c_{1}\lambda ^{3}+c_{2}\lambda
^{2}+c_{3}\lambda +c_{4},  \label{polinom}
\end{equation}%
where%
\begin{align*}
c_{1}& =2\left( 1+\gamma \right) , \\
c_{2}& =4\gamma +1+\left\vert Y\right\vert ^{2}\left( \eta ^{2}\left\vert
Y\right\vert ^{2}-4\gamma \right) , \\
c_{3}& =2\gamma \left( 1-2\left\vert Y\right\vert ^{2}\left( 1+\gamma -\eta
^{2}\left\vert Y\right\vert ^{2}\right) \right) , \\
c_{4}& =4\gamma ^{2}\left\vert Y\right\vert ^{2}\left( -1+\left\vert
Y\right\vert ^{2}\left( 1+\eta ^{2}\right) \right) ,
\end{align*}%
with $\left\vert Y\right\vert ^{2}$ given by Eq. (\ref{inthom}). Since we
are interested in the existence of dynamical behaviour, we search for a pair
of complex conjugate eigenvalues, which denote the occurrence of a Hopf
(oscillatory) instability. Following the Hurwitz criterion (see, e.g. \cite%
{Haken83}), the condition for Hopf instability $\lambda =\pm i\omega $ is
given by $c_{3}(c_{1}c_{2}-c_{3})-c_{1}^{2}c_{4}=0,$ or%
\begin{align}
0& =4\eta ^{4}\left( \gamma -1\right) \left\vert Y\right\vert ^{8}+2\eta
^{2}\left( \gamma +1\right) ^{2}\left\vert Y\right\vert ^{6}+
\label{inthopf} \\
& \eta ^{2}\left( 4\gamma ^{3}-3\gamma -3\right) \left\vert Y\right\vert
^{4}+  \notag \\
& 2\left( \gamma +1\right) ^{2}\left( 2\gamma +1\right) \left\vert
Y\right\vert ^{2}-\left( 2\gamma +1\right) ^{2}.  \notag
\end{align}

The frequency of the oscillations at the bifurcation point, given by the
imaginary part of the eigenvalue, is found by substituting $\lambda =i\omega 
$ in the polynomial. In terms of the acoustic intensity given by the
solutions of (\ref{inthopf}) the frequency reads 
\begin{equation}
\omega ^{2}=\frac{\gamma }{\gamma +1}\left( 1+2\eta ^{2}\left\vert
Y\right\vert ^{4}-2\left\vert Y\right\vert ^{2}\left( \gamma +1\right)
\right) .  \label{frec}
\end{equation}

In Fig.1 the domain of existence of Hopf bifurcations (the solutions of Eq. (%
\ref{inthopf})) is represented in the plane $<\eta ,\left\vert Y\right\vert
>,$ for different values of the relative loss parameter $\gamma
=0.01,0.1,0.25$ and $0.5.$ The stationary solutions are unstable at the left
of each curve, i.e. for small values of $\eta .$ From Fig. 1 also follows
that, for a given value of the relative loss parameter $\gamma $, a maximum
value of $\eta $ is required for the existence of dynamic solutions.

\begin{figure}[h]
\centering\includegraphics[width=8cm]{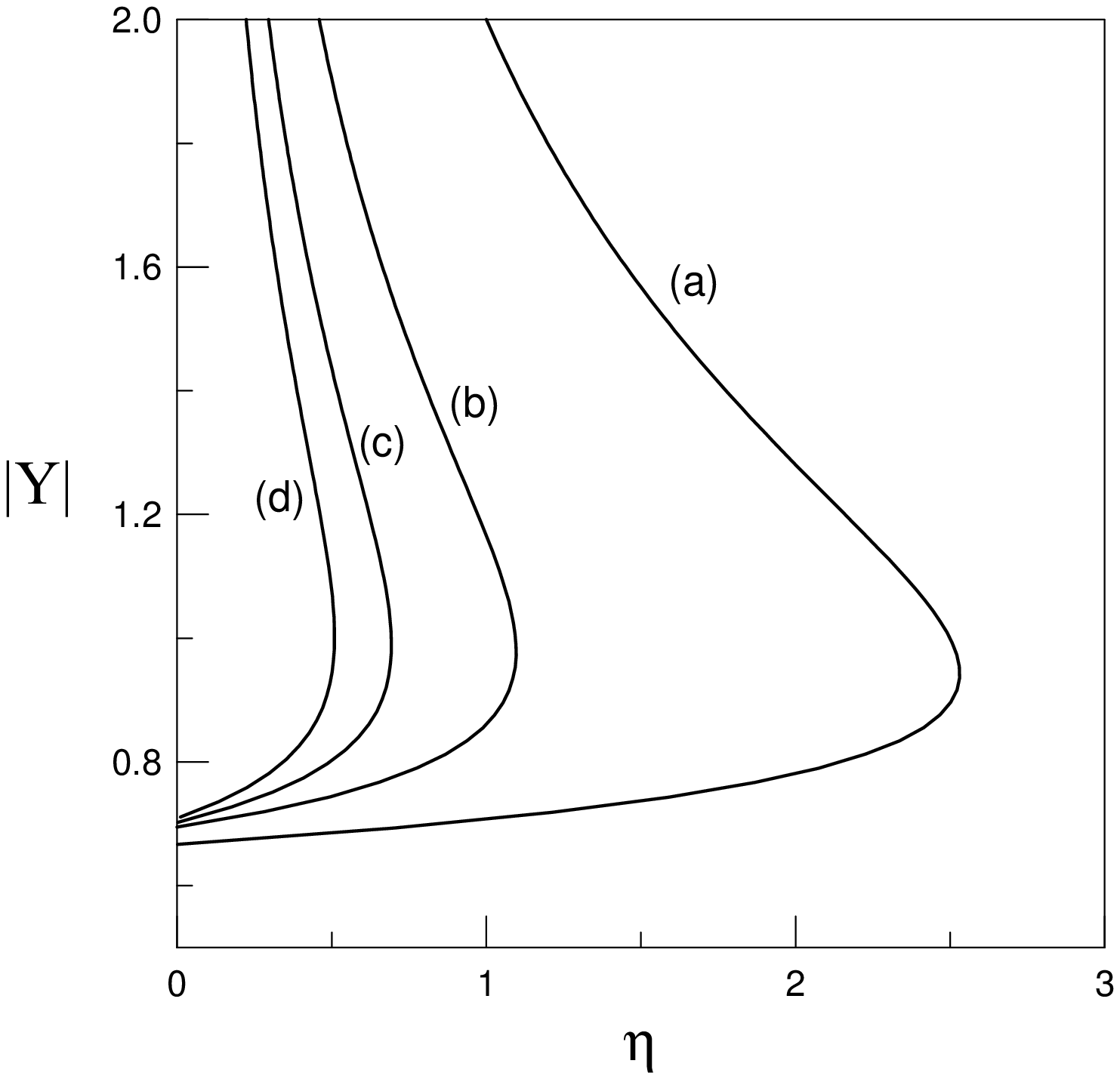}
\caption{Domain of existence of oscillatory solutions in the plane $<\protect%
\eta ,|Y|>$, for different loss parameters. Stationary solutions are
unstable at the left of each curve.}
\end{figure}

\section{Self--pulsing dynamics near homoclinic bifurcations}

In this section the numerical integration of Eqs. (\ref{eqs2}) is performed
in order to demonstrate the existence of dynamical solutions. For typical
experimental conditions, $\gamma _{Q}>\gamma _{U}$, and consequently $\gamma
<1.$ We take, according to \cite{Brysev200}, the value $\gamma =0.1$ for
numerical integration, although we note that similar results are obtained
for a smaller loss ratio. For this value, the analysis of the previous
section (see also Fig. 1, line $c$) shows that the solutions display
temporal dynamics when the nonlinearity coefficient take values in the range 
$0<\eta <0.82.$

\begin{figure}[h]
\centering\includegraphics[width=9cm]{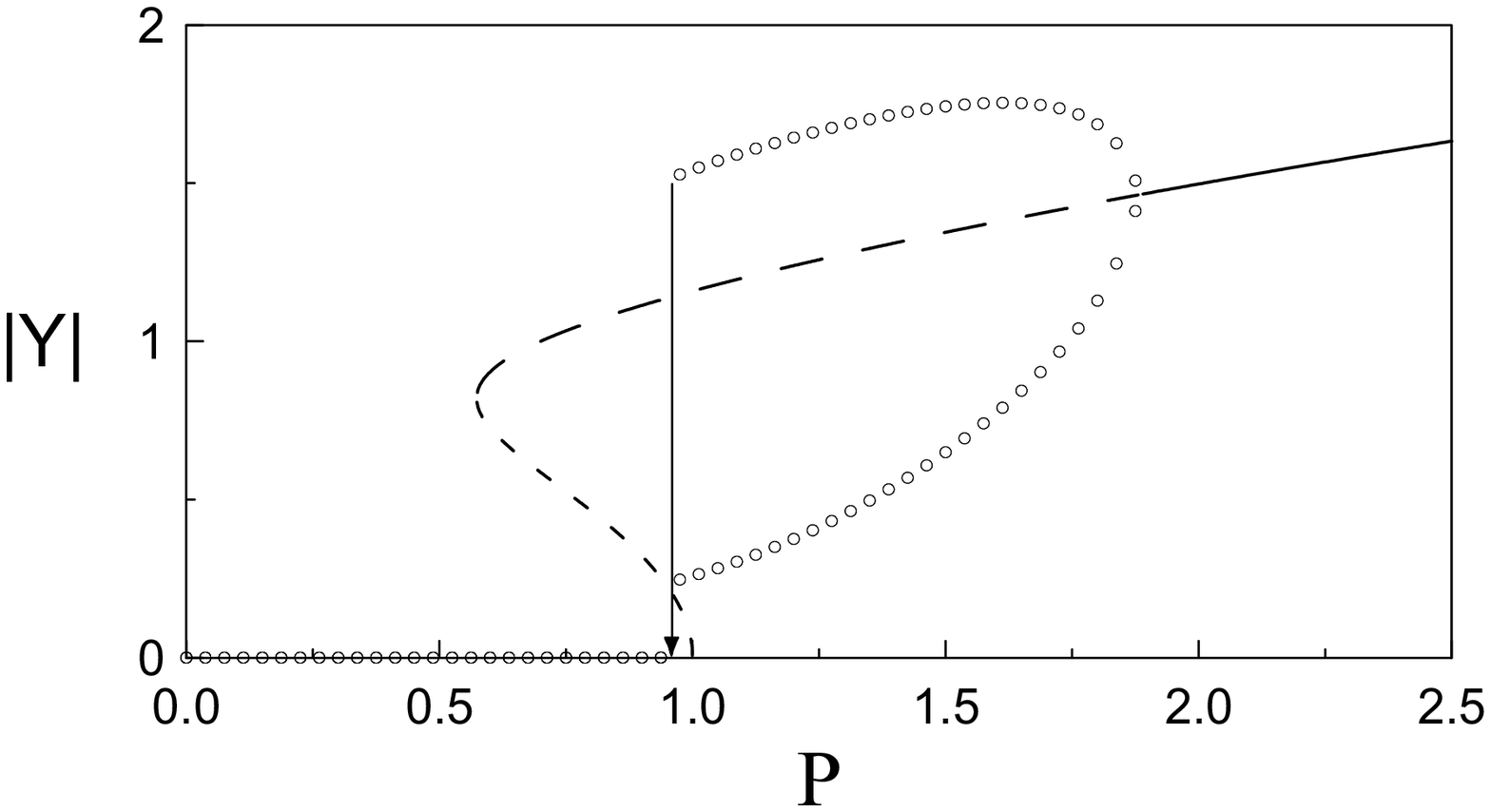}
\caption{Bifurcation diagram of the ultrasonic field for $\protect\gamma =0.1
$ and $\protect\eta =0.7$. Open circles represent maximum and minimum
amplitudes of oscillatory solutions.}
\end{figure}

In Fig. 2 the bifurcation diagram as computed numerically for $\gamma =0.1$
and $\eta =0.7$ is shown. Dashed lines correspond to the analytical
solutions given by Eq. (\ref{inthom}), and the existence of the backward
(inverted) Hopf bifurcation of the upper branch predicted by the theory of
the previous section is observed at $\mathcal{P}\approx 1.87$. Dynamical
states exist for pump values below this critical point (see also Fig. 1).
For higher pump values, the solution is stationary, and the dynamics decay
to a fixed point. The upper and lower branches with open circles in Fig. 2
correspond to maximum and minimum values of ultrasonic amplitudes in the
oscillating regime. Numerics also show the switch--off of the ultrasonic
field when pump is decreased down to a given critical value$.$\ This means
that, despite bistability, the trivial (zero) solution is a globally
attracting solution, which have important consequences on the dynamics of
the system.

Similar diagrams are obtained for any value of $\eta $ in the range given
above. For small values of the nonlinearity parameter $\eta $, the solutions
are quasi-harmonic in time regardless the pump value. However, for larger
values of $\eta $ close to limiting value $0.82$ required for the existence
of dynamical states, the solutions show a qualitatively different temporal
behaviour, not predicted by the linear stability analysis which in fact is
valid only near the bifurcation point. Next we report the numerical results
for the particular value $\eta =0.7$ corresponding to Fig. 2.

\begin{figure}[h]
\centering\includegraphics[width=9cm]{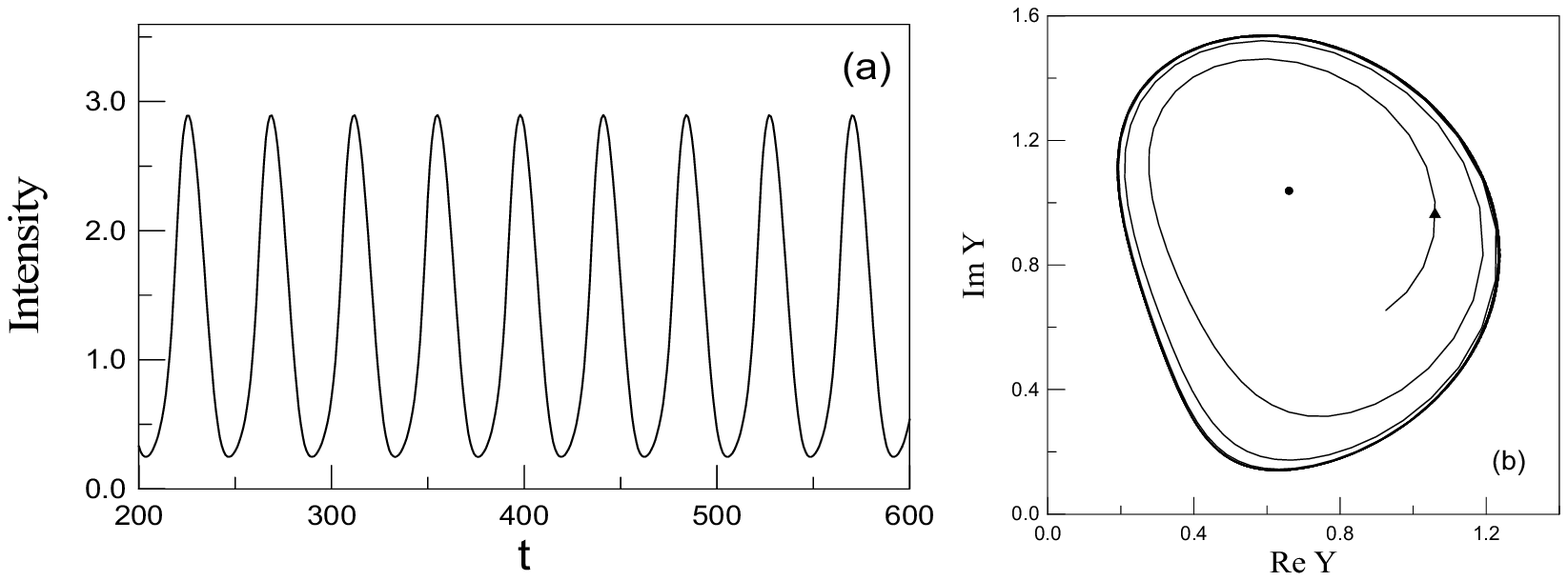}
\caption{(a) Time dependence of the intensity of the ultrasonic field for $%
\mathcal{P}=1.35$. The rest of parameters as in Fig. 2. (b) The
corresponding phase portrait.}
\end{figure}

In Fig. 3(a) the amplitude of the ultrasonic field is plotted as a funcion
of time for a pump value of $\mathcal{P}=1.35$ far from the bifurcation
point at $\mathcal{P}_{th}=1$. The corresponding phase portrait is shown in
Fig. 3(b), where the existence of a stable limit cycle is observed. However,
when decresing the pump, the ultrasonic field is emmited in the form of
periodic pulses or spikes, whose temporal separation (period) depends on the
value of the pump amplitude, and increases with it. Close to the emission
threshold the time between pulses (interspike period) tends to infinity. An
example of the temporal profile of the field amplitude in the spiking
regime, and the corresponding phase portraits for the pump value $\mathcal{P}%
=0.95$ is shown in Fig. 4(a) and (b) respectively.

\begin{figure}[h]
\centering\includegraphics[width=9cm]{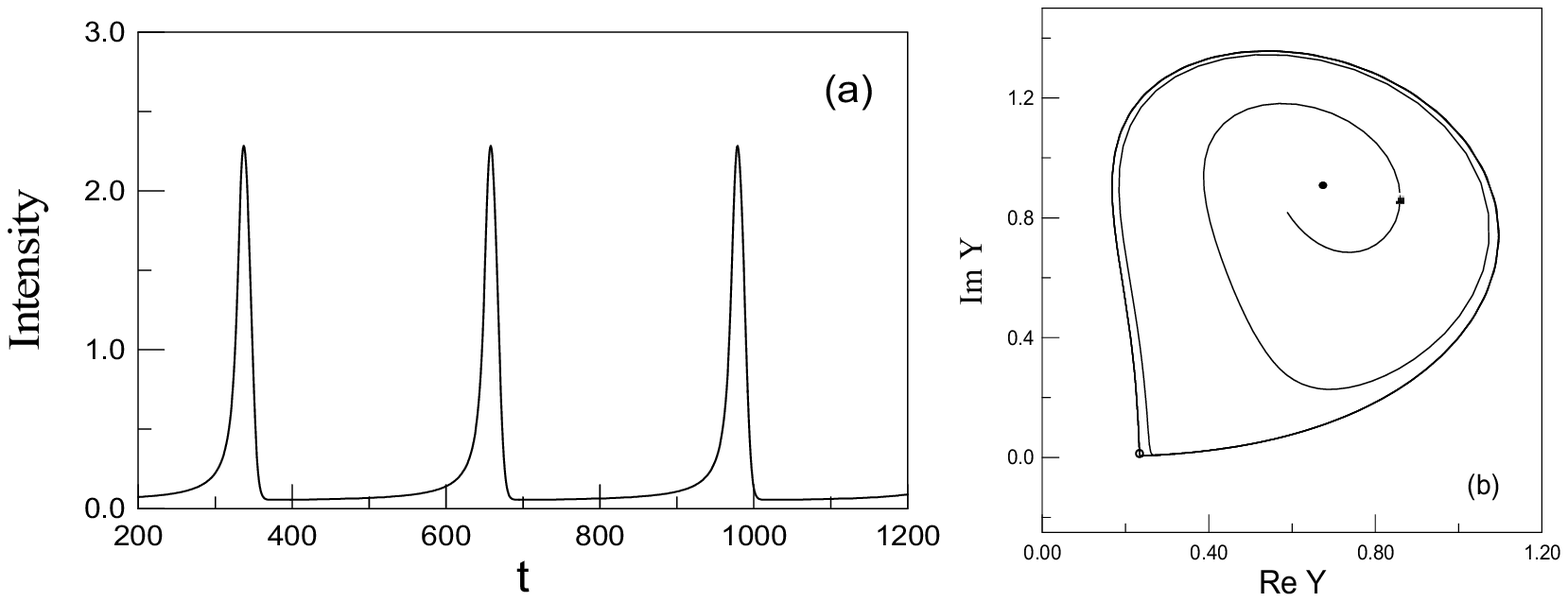}
\caption{(a) Time dependence of the intensity of the ultrasonic field for $%
\mathcal{P}=0.95$. The rest of parameters as in Fig. 2. (b) The
corresponding phase portrait. The homoclinic point is represented by the
open circle.}
\end{figure}

This behavior denotes the existence of a global bifuraction of the solutions
for the parameters considered. As the pump approaches the critical value $%
\mathcal{P}_{c}=0.904$, the amplitude of the lower branch of the oscillating
solutions (open circles in Fig. 2) approaches the lower branch of the
stationary solutions (short-dashed line in Fig. 2), which as follows from
the linear stability analysis corresponds to a saddle point. The limit cycle
at this point connects with the saddle point, and denegerates in a
saddle-loop or homoclinic bifurcation. For pump values below $\mathcal{P}%
_{c} $ the trayectory decays to a fixed point corresponding to the trivial
solution.

The period of the limit cycle near a saddle-loop bifurcation is governed by
a characteristic scaling law. Linearization of the dynamics around the
saddle leads to the following expression for the period \cite{Gaspard90} 
\begin{equation*}
T\propto -\frac{1}{\lambda }\ln (\mathcal{P}_{c}-\mathcal{P})
\end{equation*}%
where $\mathcal{P}_{c}-\mathcal{P}$ measures the distance to the homoclinic
bifurcation (which is assumed small) and $\lambda $ is the eigenvalue in the
unstable direction of the saddle point.

Figure 5 shows the numerically evaluated period of the self-pulsed solutions
in the whole range where dynamical states exists. Note the divergence in the
period at $\mathcal{P}_{c}=0.904.$ In order to check the homoclinic
character of the bifurcation, the inset shows the linear fit of the period
against $\ln (\mathcal{P}_{c}-\mathcal{P})$ for pump values close to $%
\mathcal{P}_{c}.$ The slope of the linear fit is found to be $51.4$, in good
agreement with the linear stability result $1/\lambda =54.2,$ demonstrating
the existence of the saddle-loop bifurcation.

\begin{figure}[h]
\centering\includegraphics[width=9cm]{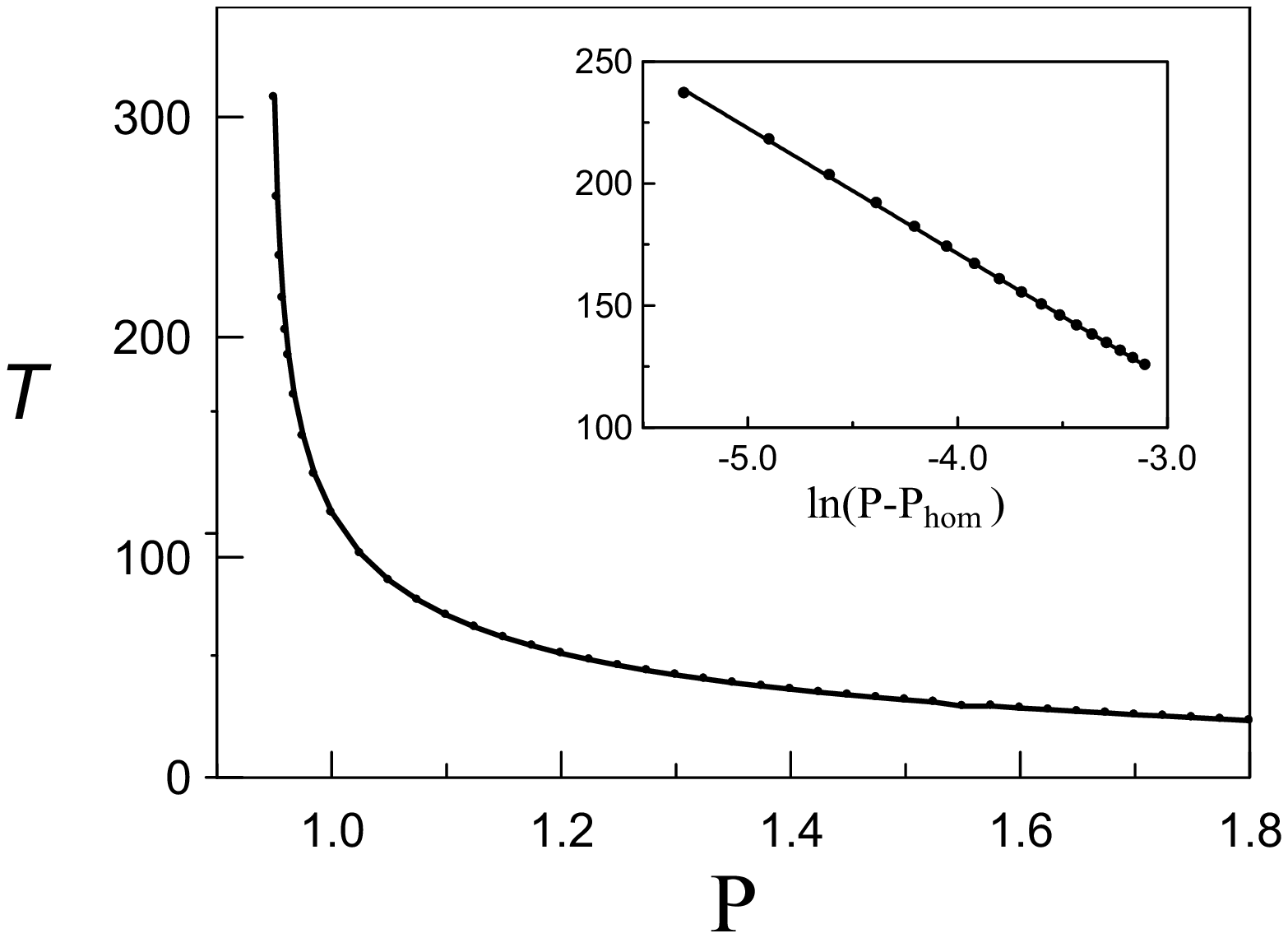}
\caption{Period of the oscillations as a function of the pump. The insert
shows the linear fit. }
\end{figure}

Finally, when decreasing the pump below the saddle--loop bifurcation point,
the subharmonic field is switched-off. Trivial solution acts then as a
globally attracting point.

\section{Conclusion}

A model of parametric generation of ultrasound in a ferromagnetic material
has been considered, and its stability analyzed, revealing the existence of
a Hopf bifurcation leading to self--pulsing dynamics. For selected values of
the decay rates and the nonlinearity parameter, the system also show a
saddle--loop bifurcation which results in a spiking regime in the emitted
ultrasound, where the frequency can take arbitrary low frequencies or firing
rates. It has been demonstrated \cite{Izhikevich00} that dynamical systems
where saddle--loop bifurcations and stable fixed points coexist have also a
property called excitability, characterized by \cite{Murray} (a)
perturbations of the rest state beyond a certain threshold induce a large
response before coming back to the rest state, and (b) there exist a
refractory time during which no further excitation is possible. This
properties, which are characteristic in several biological problems (e.g.
the behaviour of action potentials in neurons \cite{Hodking52}) and some
laser models \cite{Krauskopf03}, are being investigated in the context of
the acoustical system proposed in this paper.

\section*{Acknowledgment}

The work was financially supported by the CICYT of the Spanish Government,
under the\ project BFM2002-04369-C04-04.

\end{document}